\documentstyle[12pt]{article}
\def\tilde{\widetilde}
\def\t{\widetilde}
\def\ta{U^{1}}
\def\tb{U^{2}}
\def\ga{G^{1}_{+}}
\def\gb{G^{2}_{+}}
\def\gaa{G^{1}_{-}}
\def\gbb{G^{2}_{-}}
\def\La{L^{1}_{+}}
\def\Lb{L^{2}_{+}}
\def\laa{L^{1}_{-}}
\def\lbb{L^{2}_{-}}
\def\Lc{L^{1}_{\pm}}

\def\mc{\tilde L^{1}_{\pm}}
\def\r{r_{+}}
\def\ra{r_{-}}
\def\rb{r_{\pm}}

\def\tr{\,{\rm tr}\,}
\def\a{\alpha}
\def\b{\beta}

\def\l{\lambda}
\def\c{Chern-Simons  }
\def\G.{Gauss-law constraints. }
\def\G{Gauss-law constraints }

\def\be{\begin{equation}}
\def\ee{\end{equation}}
\def\bea{\begin{eqnarray}}
\def\eea{\end{eqnarray}}
\def\Y{Yang-Mills }

\def\a{\alpha}
\def\HU{the Heisenberg double }
\def\aah{a Heisenberg double }
\def\Hs{Heisenberg doubles }
\def\H.{the Heisenberg double. }
\def\ah.{a Heisenberg double. }
\def\Hs.{Heisenberg doubles. }
\def\g{\gamma}
\def\cb{the cotangent bundle }
\vspace{5cm}
\title{  \hfill{LMU-TPW 95-17} \\
\hfill{hep-th/9511...}\\
\vspace{1cm}
Physical phase space of lattice Yang-Mills theory and the moduli space of
flat connections on a Riemann surface}
\vspace{.7cm}
\author{ \mbox{}\\
S.A.Frolov\thanks{Alexander von Humboldt fellow} \mbox{} \\
\vspace{0.4cm}
Section Physik, Munich University
\vspace{-0.5cm} \mbox{} \\
Theresienstr.37, 80333 Munich, Germany
\thanks{Permanent address:\ Steklov Mathematical Institute, Vavilov st.42,
 GSP-1, 117966 Moscow, RUSSIA}
\mbox{}}
\date{}
\begin{document}
\maketitle
\vspace{3.5cm}
\begin{abstract}
It is shown that the physical phase space of $\g$-deformed Hamiltonian
lattice \Y theory, which was recently proposed in refs.[1,2], coincides
as a Poisson manifold with the moduli space of flat connections on a
Riemann surface with $(L-V+1)$ handles and therefore with the
physical phase space of the corresponding $(2+1)$-dimensional
Chern-Simons model, where $L$ and $V$ are correspondingly a total number of
links and vertices of the lattice. The deformation parameter $\g$ is
identified with $\frac {2\pi}{k}$ and $k$ is an integer entering the
Chern-Simons action.
\end{abstract}

\section{Introduction}

It is well-known that there are two closely-related ways to introduce
lattice regularization of gauge models. In the approach of Wilson \cite{w}
one discretizes all space-time and, thus, replaces the \Y theory by some
statistical mechanics model. In the Hamiltonian approach of Kogut and
Susskind  \cite{ks} one considers the theory in the Minkowskian space-time
and discretizes only space directions remaining the time continuous. Then
one should place on each link of the lattice some phase space and attach
to each vertex lattice Gauss-law constraints which are first-class
constraints and generate gauge transformations. Thus in the Hamiltonian
approach the continuous \Y theory is replaced by some classical mechanics
model with first-class constraints. In fact it is not difficult to show
that these two approaches are equivalent if one chooses the cotangent
bundle of a Lie group as the phase space placed on a link. However one can
consider not only the cotangent bundle and in this case the Hamiltonian
approach will lead to results which can not be derived from the Wilson
formulation.

In refs.[1,2] I have proposed  Hamiltonian lattice gauge
models based on the assignment of \aah  $D_+^{\g}$
\cite{d1,s1,sr,af1,s2,am} of a Lie group to each link. The Heisenberg
double $D_+^{\g}$ depends on one complex parameter $\g$ and the cotangent
bundle of a Lie group can be regarded as a limiting case of $D_+^{\g}$
when $\g$ goes to zero.  Quantization of \HU leads to an algebra which
contains as subalgebras the quantized universal enveloping algebra
$U_{q}(\cal G)$ and the algebra of functions on the quantum group
$Fun_{q}(G)$ \cite{af1,s2,af2}, and $q$ is related to $\g$ by means of the
following formula: $q={\rm e}^{i\hbar\g}$.  Only the case of imaginary
$\g$ (real $q$) was considered in refs.[1,2].  In the present paper we
find a proper generalization to the case of real $\g$ ($q$ lying on the
unit circle). This case seems to be of the most importance in quantum
theory since for $q$ a root of unity \HU has just a finite number of
irreducible finite-dimensional representations and thus the Hilbert space
is finite-dimensional too. It permits to develop the weak-coupling
expansion for the Hamiltonian lattice \Y theory which differs from the
standard perturbation theory.

The simplest way to get such a generalization seems to be to study the
structure of the physical phase space of the usual Hamiltonian lattice
gauge models on graphs, and then to deform the physical phase space. We
show that the gauge invariance of a gauge model on an arbitrary lattice
(or a graph) can be used to reduce the graph to a standard graph with one
vertex and $g=L-V+1$ loops (tadpoles), where $L$ and $V$ are a total
number of links and vertices of the original graph. The Gauss-law
constraints attached to the only vertex of the graph generate residual
gauge transformations on the reduced phase space, which is just the direct
product of cotangent bundles over all tadpoles. Generalization of gauge
models on standard graphs can be obtained in the same way as was done in
ref.[2] by replacing the cotangent bundle by \HU and the residual Gauss-law
constraints by first-class constraints generating the well-known dressing
transformations \cite{s1,s2} . We note that the Poisson algebra obtained
coincides after some transformation \cite{AM,A} with the Poisson algebra
introduced by Fock and Rosly \cite{fock} to describe the Poisson structure
of the moduli space of flat $SL(N)$ connections on a Riemann surface with
$g=L-V+1$ handles and find a new antiautomorphism of the Poisson algebra
which permits to single out the moduli space of flat $SU(N)$ connections.
Thus the $\g$-deformed Hamiltonian lattice \Y theory and (2+1)-dimensional
Chern-Simons theory with $k=\frac {2\pi}{\g}$ have the same physical phase
space. Due to the well-known result of Witten \cite{wit,ww} the Hilbert
space of the (2+1)-dimensional Chern-Simons theory and, therefore, of the
lattice \Y theory is finite-dimensional and coincides with the space of
conformal blocks of the WZNW model.

The plan of the paper is as follows. In the second section we consider
gauge models on arbitrary graphs and the procedure of reduction to a
standard graph. In the third section we firstly remind some simple results
from the theory of \H. Then the deformation of the physical phase space
of the Hamiltonian lattice \Y theory is described and the relation to the
moduli space of flat connections and to the (2+1)-dimensional Chern-Simons
theory is pointed out. In Conclusion we discuss unsolved problems and
perspectives.

\section{Gauge models on graphs}

In this section we firstly consider gauge models  on   arbitrary graphs
(regular hyper-cubic lattice, triangulation of a surface,
simplicial complexes and so on) and then we show that any gauge model on a
graph can be reduced to a gauge model on a standard graph.  Any graph is
described by a set of vertices and a set of links.  Each link is thought
of as either a path connecting two vertices or a closed path with a marked
vertex (tadpole).  Two vertices can be connected by any finite number of
links.  Such a graph is certainly just an arbitrary connected Feynman
diagram.

Let us now consider some vicinity of a vertex $v$ which does
not contain other vertices and closed paths. Let us denote the paths
which go from the vertex $v$ by $l_{1}(v)$,...,$l_{N_{v}}(v)$. We call
such a path as a vertex path. $N_{v}$ is a common number of the paths and
if there is no closed path for the vertex $v$ then $N_{v}$ coincides
with the number of links going from $v$ to some other vertices of the
lattice. With each vertex path $l_{i}(v)$ one associates a field taking
values in \cb $T^{*}G$ of a Lie group. This field can be described by
a group-valued matrix $U(l_{i}(v))$, and an algebra-valued
matrix $E(l_{i}(v))$ with the standard Poisson structure
\bea
\{ U^{1},U^{2}\} &=&0 \nonumber \\
\{ E^{1},E^{2}\} &=&\frac {1}{2} [E^{1}-E^{2},C] \nonumber \\
\{ E^{1},U^{2}\} &=&CU^{2}
\label{2.1}
\eea
and fields corresponding to different paths have vanishing Poisson brackets.

\noindent In eq.(2.1) we use the standard notations from the theory of
quantum groups \cite{d2,frt}:  for any matrix $A$ acting in some space $V$
one can construct two matrices $A^{1}=A\otimes id $ and $A^{2}=id\otimes A $
acting in the space $V\otimes V$, the matrix $C$ is the tensor Casimir
operator of the Lie algebra ${\cal G}$ of the group $G$:
$C=-\eta_{ab}\l^{a}\otimes\l^{b}$ and $\eta_{ab}$ is the Killing tensor and
 $\l^{a}$ form a basis of ${\cal G}$.

 One can see from eq.(2.1) that the field $E$ should be identified with the
 right-invariant momentum generating left gauge transformations of the
 field $U$. It is useful to introduce a different parametrization of
 $T^{*}G$ by means of the left-invariant momentum $\t E=-U^{-1}EU$ and of
 the matrix $\t U=U^{-1}$. One can easily check that the fields $\t U$ and
 $\t E$ have the same Poisson structure (2.1), momenta $E$ and $\t E$ have
 vanishing Poisson bracket and we shall need the following expression for
 the bracket of $\t E$ and $U$
 \be
 \{\t E^{1},U^{2}\}=-U^{2}C
 \label{2.2}
 \ee

Let us now attach to the vertex $v$ the following
Gauss-law constraints
\be
G(v)= \sum_{i=1}^{N_{v}} E(l_{i}(v))=0
\label{2.3}
\ee
These constraints form the Poisson-Lie algebra
\be
\{ G^{1},G^{2}\} =\frac {1}{2} [G^{1}-G^{2},C]
\label{2.4}
\ee
 and generate the following gauge transformations of the fields
$U(l_{i}(v))$ and $E(l_{i}(v))$
\bea
U(l_{i}(v))&\to& g(v)U(l_{i}(v)) \nonumber \\
E(l_{i}(v))&\to& g(v)E(l_{i}(v))g^{-1}(v)
\label{2.5}
\eea

Repeating the same procedure for all of the vertices one gets the phase space
which is the direct product of cotangent bundles over all of the
vertex paths and a set of the \G attached to the vertices. The \G
corresponding to different vertices have vanishing Poisson brackets.
Taking into account that for each link there are two vertex paths one sees
that one has placed on each link two different cotangent bundles. However
one can identify these bundles using the fact that the fields $U$ and $E$
 and $\t U$ and $\t E$ have the same Poisson structure. Thus one can
impose on the fields attached to one link the following constraints
\bea
\t U(1)&=&U^{-1}(1)=U(2) \nonumber \\
\t E(1)&=&-U^{-1}(1)E(1)U(1)=E(2)
\label{2.6}
\eea
So the fields $U(1)$, $E(1)$ and $U(2)$, $E(2)$ are just different
coordinates on the same cotangent bundle. The final phase space of the
model is thus the direct product of cotangent bundles over all links:
$\prod_{links} T^*G$.

Let us note that due to the constraints (2.6) the field
$U(l(v_{1},v_{2}))$ corresponding to a link $l(v_{1},v_{2})$ which
 connects vertices $v_{1}$ and $v_{2}$ is transformed by the \G $G(v_{1})$
 and $G(v_{2})$ as follows
\be
U(l(v_{1},v_{2}))\to g(v_{1})U(l(v_{1},v_{2}))g^{-1}(v_{2})
\label{2.7}
\ee
This is the usual transformation law in lattice \Y theory. However the
field $U(l(v))$ corresponding to a tadpole $l(v)$ attached to a vertex $v$
is transformed by means of conjugations
\be
U(l(v))\to g(v)U(l(v))g^{-1}(v)
\label{2.8}
\ee
We see from eq.(2.8) that one can not eliminate the field $U(l(v))$ by
means of a gauge transformation.

Observables which are invariant with respect to the gauge transformations
(2.7) and (2.8) can be constructed in a standard way. If the graph under
consideration is a regular hyper-cubic lattice one gets the usual
lattice \Y model with the following Hamiltonian (which is certainly not
unique)
\be
H=-\frac {e^{2}}{2}a^{2-d} \sum_{links}\tr E^{2}(l) -
\frac {a^{d-4}}{8e^{2}}\sum_{plaquettes}\big( W(\Box) + W^{*}(\Box)\big)
\label{2.9}
\ee
Here the summation is taken over all links and over all
plaquettes, $d$ is a dimension of space, $e$ is a coupling constant,
$a$ is a lattice length and $W(\Box)$ is the usual Wilson term.

Let us now introduce the notion of the gauge equivalence of two graphs.
Two graphs are called gauge equivalent if the corresponding gauge models
have the same physical phase space. Let us remind that the physical phase
space can be obtained by imposing some gauge conditions and then by
solving the \G. We shall show that any graph is equivalent to a standard
graph with one vertex and $g=L-V+1$ links (all links are tadpoles), where
$L$ and $V$ are correspondingly a total number of links and vertices of
the original graph.

To prove the statement let us consider some link $l$ connecting two
different vertices $v_{1}$ and $v_{2}$. There are two vertex paths
$l(v_{1})$ and $l(v_{2})$ corresponding to the link $l$. In what follows
we denote the vertex path $l(v_{1})$ as $l$ and $l(v_{2})$ as $l^{-1}$.
The constraints (2.6) imply $U(l^{-1})=U^{-1}(l)$ and
$E(l^{-1})=-U^{-1}(l)E(l)U(l)$. Using the gauge invariance under the
transformation (2.7) one can impose the gauge condition $U(l)=1$. Then one
has to express the corresponding momentum $E(l)$ through the remaining
variables of the phase space. The field $E(l)$ enters two \G $G(v_{1})$ and
$G(v_{2})$ as follows
\be
G(v_{1})=E(l)+ \sum_{\hbox{paths}} ' E(l_{i}(v_{1}))=0
\label{2.10}
\ee
\bea
G(v_{2})&=&E(l^{-1})+ \sum_{\hbox{paths}} ' E(l_{i}(v_{2}))= \nonumber\\
&=&-E(l)+ \sum_{\hbox{paths}} ' E(l_{i}(v_{2}))=0
\label{2.11}
\eea
where the summation in eqs.(2.10) and (2.11) goes over all vertex paths
excepting $l(v_{1})$ and $l(v_{2})$ correspondingly.

\noindent One can find the field $E(l)$ from eq.(2.10) and inserting the
solution into eq.(2.11) one gets instead of two constraints $G(v_{1})$ and
$G(v_{2})$ a residual constraint
\be
G(v_{1},v_{2})=\sum_{\hbox{paths}} ' E(l_{i}(v_{1}))+\sum_{\hbox{paths}}
' E(l_{i}(v_{2}))=0
\label{2.12}
\ee
Now it remains to note that the
same \G correspond to a graph which is obtained from the original graph by
shrinking the link $l$ and thus by identifying the vertices $v_{1}$ and
$v_{2}$. Proceeding in the same way one finally gets the standard graph
with one vertex and one residual constraint which has the following form
\be
G= \sum_{i=1}^{g} E(i)+\t E(i)=\sum_{i=1}^{g} E(i)-U^{-1}(i)E(i)U(i)=0
\label{2.13}
\ee
where $g=L-V+1$ is the number of links of the standard graph.

\noindent It is not difficult to check that the residual gauge
transformations generated by this constraint are the simultaneous
conjugations
\be
U(i)\to gU(i)g^{-1},\qquad  E(i)\to gE(i)g^{-1}
\label{2.14}
\ee
It is clear that the gauge fixing just described corresponds to a choice
of a maximal tree on a graph. The physical phase space can be now obtained
as a factor space of the space, which is the result of the
solution of the constraint (2.13), over  the action (2.14) of the residual
gauge group. This phase space is not a manifold because the gauge group
action is not free. This fact seems to be closely related to the
well-known Gribov ambiguity.

The reduction procedure just described can be used to calculate the
reduced Hamiltonian. In particular it is possible to show that the
magnetic part of the Hamiltonian of the (2+1)-dimensional lattice \Y
theory defined on a square lattice with free boundary conditions can be
reduced to the following form
\be
H_{m}=\frac {1}{e^{2}a^{2}}\sum_{i=1}^{g} tr \big( U^{-1}_{i}+U_{i}\big)
\label{2.15}
\ee
and $g$ is equal to the number of plaquettes in this case.

Unfortunately the spectrum of this Hamiltonian is continuous and one can
not use it to develop the weak coupling expansion. However let us suppose
that we have a way to compactify the physical phase space. It is known
that quantization of a compact phase space leads to a finite-dimensional
Hilbert space and, therefore, any operator acting in the space has a
discrete spectrum and one can easily apply standard perturbation theory.
The compactification can be achieved by replacing the cotangent bundles by
the Heisenberg doubles $D_{+}^{\g}$ and will be discussed in the next
section. Let us finally note that in the (2+1)-dimensional case there is
another and, may be, more attractive possibility to develop the weak
coupling expansion. Taking into account that the electric part of the
reduced Hamiltonian contains a term which is proportional to
\be
H_{e}^{0}=e^{2}\sum_{i=1}^{g} tr E_{i}^{2}
\label{2.16}
\ee
one can use the sum of the Hamiltonian $H_{m}$ and $H_{e}^{0}$ as the
first approximation. For $SU(2)$ group the Hamiltonian
$H^{0}=H_{m}+H_{e}^{0}$ describes an  exactly-solvable model and one may
hope to calculate exactly its spectrum. It is worthwhile to note that the
same Hamiltonian describes the superfluid B-phase of $^{3}$He. It
would be very interesting to find an integrable generalization of the
Hamiltonian $H^{0}$ for the $SU(N)$ group.

\section{Deformation of the physical phase space}

In this section we firstly remind some simple results from the theory of
\HU (for detailed discussion see refs.\cite{d1,s1,sr,af1,s2,am}). Then we
deform the physical phase space of gauge models on a graph by replacing
the cotangent bundle and the residual Gauss-law constraint (2.13) by \HU
and by some deformed constraint correspondingly. After that we show that
the deformed phase space coincides as a Poisson manifold with the moduli
space of flat connections on a Riemann surface with $g=L-V+1$ handles.

Let $G$ be a matrix algebraic group and $D=G\times G$. For definiteness we
consider the case of the $SL(N)$ group. Almost all elements $(x,y) \in D$
can be presented in two equivalent forms  as follows
\bea
(x,y)&=&(U,U)^{-1}(L_{+},L_{-})=(U^{-1}L_{+},U^{-1}L_{-}) \nonumber\\
&=&(\tilde L_{+}, \tilde L_{-})^{-1}(\tilde U,\tilde U)=
(\tilde L_{+}^{-1} \tilde U,\tilde L_{-}^{-1}\tilde U)
\label{3.1}
\eea
where $U, \tilde U \in G$, the matrices $L_{+}, \tilde L_{+}$ and
$L_{-},\tilde L_{-} $ are upper- and lower-triangular, their diagonal parts
$l_{+}, \tilde l_{+}$ and $l_{-},\tilde l_{-}$ being inverse to each other:
$l_{+}l_{-}= \tilde l_{+}\tilde l_{-}=1$.

Let all of the matrices be in the fundamental representation  $V$ of the group
$G$ ($N\times N$ matrices for the $SL(N)$ group). Then the algebra of
functions on the group $D$ is generated by  the matrix elements $x_{ij}$ and
$y_{ij}$. The matrices $L_{\pm}$ and $U$ or $\tilde L_{\pm}$ and  $\tilde U$
can be considered as almost everywhere regular functions of $x$ and $y$.
Therefore, the  matrix elements $L_{\pm ij}$ and $U_{ij}$
(or $\tilde L_{\pm ij}$ and $\tilde U_{ij}$) define another system of
generators of the algebra $Fun D$. We define the Poisson structure on the
group $D$ in terms of the generators $L_{\pm}$ and $U$ as follows
\cite{af1,s2}
\be
\{\ta ,\tb \} =\g [\rb ,\ta\tb ]
\label{3.2}
\ee
\bea
&&\{\La ,\Lb \} =\g [\rb ,\La\Lb ] \nonumber\\
&&\{\laa ,\lbb \} =\g [\rb ,\laa\lbb] \nonumber\\
&&\{\La ,\lbb \} =\g [\r ,\La\lbb]
\label{3.3}
\eea
\bea
&&\{\La ,\tb \} =\g \r \La\tb \nonumber\\
&&\{\laa ,\tb\} =\g \ra \laa\tb
\label{3.4}
\eea
Here $\g$ is an arbitrary complex parameter, $\rb$ are classical $r$-matrices
which satisfy the classical Yang-Baxter equation and the following relations
\be
[r^{12} ,r^{13} ]+[r^{12} ,r^{23} ]+[r^{13} ,r^{23} ]=0
\label{3.5}
\ee
\be
\ra =-P\r P, \qquad \r -\ra =C
\label{3.6}
\ee
where $P$ is a permutation in the tensor product $V\otimes V$ ($Pa\otimes
b=b\otimes a$). For the $SL(N)$ group the solution of
eqs.(\ref{3.5}-\ref{3.6}) looks as follows
\bea
\r &=& \sum_{i=1}^{N-1}
h_{i}\otimes  h_{i} +2\sum_{i<j}^{N} e_{ij}\otimes e_{ji} \nonumber\\
&=&-\frac {1}{N} I + \sum_{i=1}^{N} e_{ii}\otimes e_{ii}  + 2\sum_{i<j}^{N}
e_{ij}\otimes e_{ji}
\label{3.7}
\eea
where $(e_{ij})_{kl} =\delta_{ik}\delta_{jl}$ and $h_{i}$ form an orthonomal
basis of the Cartan subalgebra of the $SL(N)$ group:
$\sqrt{\, i(i+1)} h_{i}=\sum_{k=1}^{i} e_{kk}-ie_{i+1,i+1}$.

\noindent In eq.(\ref{3.5}) using the matrix $r=\sum_{a} r_{1}(a)
\otimes r_{2}(a)$ acting in the space $V\otimes V$ one constructs matrices
$r^{12}=\sum_{a} r_{1}(a)\otimes r_{2}(a)\otimes id$,
$r^{13}=\sum_{a} r_{1}(a)\otimes id\otimes r_{2}(a)$ and
$r^{23}=\sum_{a} id\otimes r_{1}(a)\otimes r_{2}(a)$ acting in the space
$V\otimes V\otimes V$.

The group $D$ endowed with the Poisson structure (\ref{3.2}-\ref{3.4}) is
called the Heisenberg double $D_{+}^{\g}$ of the group $G$.  It is not
difficult to show that the matrices $\tilde L_{\pm}$ and $\tilde U$ have the
same Poisson structure (\ref{3.2}-\ref{3.4}) and we shall need the Poisson
brackets of $L_{\pm}$, $U$ and $\tilde L_{\pm}$, $\tilde U$ \cite{af2}
\bea \{L_{\a}^{1},\tilde L_{\b}^{2}\} &=&0 \qquad for \quad any \quad \a
,\b =+,- \nonumber \\ \{\mc ,\tb\} &=&-\g\mc \tb \rb \nonumber \\ \{\Lc
,\tilde\tb\} &=&-\g\Lc \tilde\tb \rb \nonumber \\ \{\ta ,\tilde\tb\} &=&0
\label{3.8}
\eea
The cotangent bundle of the group $G$ can be considered as a limiting case of
\H. Namely, in the limit $\g \to 0$ and  $L_{\pm} \to 1+\g E_{\pm}$,
$E=E_{+}-E_{-}$ the Poisson structure of \HU coincides with the canonical
Poisson structure of the cotangent bundle $T^{*}G$.

Now we are ready to discuss the deformation of  gauge models on graphs.
The case of imaginary $\g$ was considered in ref.[2] for gauge
models on arbitrary graphs. The real $\g$ case is more complicated and it
seems to be possible to get a proper deformation only for gauge models on
a standard graph. However it does not mean any loss of information because
as was shown in preceding section gauge models on arbitrary graphs are
equivalent to gauge models on standard graphs.

We begin with \HU of the complex $SL(N)$ group and discuss the equation
which singles out the real $SU(N)$ form later on. So let us place on each
link of a standard graph with $g$ links a Heisenberg double. The phase
space is thus the direct product of Heisenberg doubles over all links
$\prod_{links} D_{+}^{\g}$. Then one should replace the \G (2.13) by some
first-class constraints which reduce to the form (2.13) in the limit
$\g\to 0$. We use the following constraints
\bea
G_{\pm}&=&\tilde L_{\pm}(1)L_{\pm}(1)\tilde L_{\pm}(2)L_{\pm}(2)\cdots \t
L_{\pm}(g)L_{\pm}(g)=\nonumber\\ &=&G_{\pm}(1)G_{\pm}(2)\cdots
G_{\pm}(g)=1 \label{3.12}
\eea
and we introduced a natural notation $G_{\pm}(i)\equiv \tilde
L_{\pm}(i)L_{\pm}(i)$.  It is not difficult to verify that these
constraints have the following Poisson brackets \bea &&\{\ga ,\gb \} =\g
[\rb ,\ga\gb ]  \nonumber\\ &&\{\gaa ,\gbb \} =\g [\rb ,\gaa\gbb]
\nonumber\\ &&\{\ga ,\gbb \} =\g [\r ,\ga\gbb] \label{3.13} \eea These
Poisson brackets vanish on the constraints surface $G_{\pm}=1$ and
therefore they are first-class constraints. It is useful to consider
instead of two constraints $G_{+}$ and $G_{-}$ one constraint
$G=G_{-}^{-1}G_{+}=1$. This constraint satisfies the following quadratic
Poisson algebra
\be
\frac {1}{\g}\{ G^{1},G^{2}\} =G^{1}\r G^{2}-G^{2}G^{1}\r -
\ra G^{2}G^{1}+G^{2}\ra G^{1}
\label{3.14a}
\ee
In the limit $\g\to 0$, $L_{\pm}\to 1+\g E_{\pm}$  it reduces
to the usual Gauss-law constraint (2.13).

So we have defined the deformed phase space and \G and now one should just
remember that the same phase space and constraints recently appeared in
\cite{AM,A} where the relation between  \HU and the symplectic
structure of the moduli space of flat connections on a Riemann surface was
studied. Namely it was shown in \cite{AM,A} that there is such a change of
variables that the Poisson structure in terms of the new variables
coincides with the Poisson structure which was introduced by Fock and
Rosly \cite{fock} to describe the moduli space. For reader's convinience we
present here the corresponding formulas in our notations.

So let us consider the following change of variables \cite{AM,A}
\bea
&&A_{i}=W_{i}^{-1}L_{-}^{-1}(i)U(i)L_{+}(i)W_{i}  \nonumber\\
&&B_{i}=W_{i}^{-1}L_{-}^{-1}(i)L_{+}(i)W_{i}   \nonumber\\
&&W_{i}=G_{+}(i+1)\cdots G_{+}(g), \quad W_{g}=1
\label{3.14}
\eea
The Gauss-law constraint $G=G_{-}^{-1}G_{+}=1$ can be expressed through
the new fields $A_{i}$ and $B_{i}$ as follows
\bea
G^{-1}=G_{+}^{-1}G_{-}&=&A_{1}^{-1}B_{1}A_{1}B_{1}^{-1}
A_{2}^{-1}B_{2}A_{2}B_{2}^{-1}\cdots
A_{g}^{-1}B_{g}A_{g}B_{g}^{-1}=  \nonumber\\
&=&\prod_{i=1}^{g} A_{i}^{-1}B_{i}A_{i}B_{i}^{-1}=1
\label{3.15}
\eea
Eq.(\ref{3.15}) is the defining relation for the holonomies $A_{i}$ and
$B_{i}$ of a flat connection along the cycles $a_{i}$ and $b_{i}$ of a
Riemann surface with $g$ handles. Using the Poisson structure of  \HU
one can easily calculate the Poisson structure of the fields $A_{i}$ and
$B_{i}$ \bea &&i=1,\cdots ,g \nonumber\\ \frac {1}{\g}\{
A_{i}^{1},A_{i}^{2}\}&=&A_{i}^{1}\r A_{i}^{2}- A_{i}^{2}A_{i}^{1}\r -\ra
A_{i}^{2}A_{i}^{1}+A_{i}^{2}\ra A_{i}^{1} \nonumber\\ \frac {1}{\g}\{
B_{i}^{1},B_{i}^{2}\}&=&B_{i}^{1}\r B_{i}^{2}- B_{i}^{2}B_{i}^{1}\r -\ra
B_{i}^{2}B_{i}^{1}+B_{i}^{2}\ra B_{i}^{1} \nonumber\\ \frac {1}{\g}\{
A_{i}^{1},B_{i}^{2}\}&=&A_{i}^{1}\r B_{i}^{2}- B_{i}^{2}A_{i}^{1}\r -\r
B_{i}^{2}A_{i}^{1}+B_{i}^{2}\ra A_{i}^{1} \nonumber\\ &&i<j \nonumber\\
\frac {1}{\g}\{ A_{i}^{1},A_{j}^{2}\}&=&A_{i}^{1}\r A_{j}^{2}-
A_{j}^{2}A_{i}^{1}\r -\r A_{j}^{2}A_{i}^{1}+A_{j}^{2}\r A_{i}^{1}
\nonumber\\
\frac {1}{\g}\{ A_{i}^{1},B_{j}^{2}\}&=&A_{i}^{1}\r B_{j}^{2}-
B_{j}^{2}A_{i}^{1}\r -\r B_{j}^{2}A_{i}^{1}+B_{j}^{2}\r A_{i}^{1}
\nonumber\\
\frac {1}{\g}\{ B_{i}^{1},B_{j}^{2}\}&=&B_{i}^{1}\r B_{j}^{2}-
B_{j}^{2}B_{i}^{1}\r -\r B_{j}^{2}B_{i}^{1}+B_{j}^{2}\r B_{i}^{1}
\nonumber\\
\frac {1}{\g}\{ B_{i}^{1},A_{j}^{2}\}&=&B_{i}^{1}\r A_{j}^{2}-
A_{j}^{2}B_{i}^{1}\r -\r A_{j}^{2}B_{i}^{1}+A_{j}^{2}\r B_{i}^{1}
\label{3.16}
\eea
The Poisson structure (\ref{3.16}) coincides with the structure which was
introduced in \cite{fock} for the description of the moduli space of flat
$SL(N)$ connections on a Riemann surface with $g$ handles. The Gauss-law
constraint (\ref{3.15}) generates the following gauge transformations
\be
A_{i}\to gA_{i}g^{-1}, \quad B_{i}\to gB_{i}g^{-1}
\label{3.17}
\ee
where the gauge parameters $g$ depend on the constraint $G$.

\noindent From the point of view of the theory of Poisson-Lie groups one
should regard the gauge parameters $g$ as belonging to a Poisson-Lie
group. Then eq.(\ref{3.17}) defines an action of the Poisson-Lie group on the
Poisson algebra (\ref{3.16}) by means of so-called dressing
transformations \cite{s1,s2}.

We have considered up to now only the case of complex $SL(N)$ group.
However for physical applications one has to single out the $SU(N)$ real
form. It can be done by means of the following condition which seems to
be unknown before
\be
A^{*}_{i}=G_+ A^{-1}_{i}G_+^{-1}, \quad
B^*_{i}=G_+B^{-1}_{i}G_+^{-1}
\label{3.18}
\ee
Here $A^{*}$ is a matrix hermitian-conjugated to $A$.

\noindent It is of no problem to
check that this condition is compatible with the Poisson structure
(\ref{3.16}) and with the Gauss-law constraint (\ref{3.5}).  The
corresponding anti-automorphism
\be
\rho (A_{i})=G_+ A^{-1}_{i}G_+^{-1}, \quad
\rho (B_{i})=G_+ B^{-1}_{i}G_+^{-1}
\label{aut}
\ee
is not an anti-involution of the Poisson algebra (\ref{3.16}).

\noindent It would be interesting to compare this anti-automorphism   with
the involution  introduced in \cite{ags} to quantize the moduli space. Let
us note that on the constraints surface $G_{\pm}=1$ the condition
(\ref{3.18}) is the standard involution which singles out the $SU(N)$
group.

So we have shown that the
$\g$-deformed physical phase space of a gauge model on a graph coincides
with the moduli space of flat connections on a Riemann surface and is
compact for the $SU(N)$ group. It is well-known \cite{wit} that the same
moduli space is a physical phase space of the (2+1)-dimensional
Chern-Simons theory, the parameter $\g$ being identified with $\frac
{2\pi}{k}$. The Chern-Simons parameter $k$ is required to be integer for
the $SU(N)$ group. Due to this relation all correlation functions of the
lattice \Y theory may be expressed through nonlocal correlation functions
of the Chern-Simons theory. Quantization of the physical phase space leads
to a finite-dimensional Hilbert space which can be identified with the
space of conformal blocks of the WZNW model \cite{wit,ww} .

\section{Conclusion}

In this paper the structure of the physical phase space of gauge models on
graphs was studied. Any graph was shown to be gauge equivalent to a
standard graph and the reduction procedure to the standard graph was
described.

The deformation of gauge models on standard graphs based on the assignment
of \aah to each link was discussed. The physical phase space of the
deformed $SL(N)$ gauge model was proved to coincide with the moduli space
of flat $SL(N)$ connections on a Riemann surface and an equation
(3.35) which singles out the moduli space of flat $SU(N)$ connections was
found. As is well-known the same moduli space is the physical phase space
of the (2+1)-dimensional \c model. By this reason all correlation
functions of deformed gauge models can be expressed through nonlocal
correlation functions of the \c model.

In quantum theory the physical Hilbert space of the \c model is known
\cite{wit} to be finite-dimensional and can be identified with the space
of conformal blocks of the WZNW model. It would be interesting by using
the relation to the WZNW model to reformulate the eigenvalue problem for
the reduced \Y Hamiltonian in terms of conformal field theory.

The finite-dimensionality of the Hilbert space gives also a possibility to
develop a weak-coupling expansion which differs from the asymptotic
expansion of the standard perturbation theory. The main contribution in
the weak-coupling expansion is given by the magnetic part $H_m$ of the \Y
Hamiltonian. Although the form of the Hamiltonian $H_m$ depends on the
space dimension, it is obvious that any Hamiltonian $H_m$ describes an
integrable system and, moreover, all of them belongs to the same
integrable hierarchy. So, the first step in the weak-coupling expansion is
to solve the corresponding integrable systems. It seems to be possible to
carry out at least in two space dimensions due to the factorization (2.15)
of $H_m$.

It is worthwhile to note that such a deformation of lattice \Y theory
leads to some kind of duality between magnetic and electric fields.
The $A$ and $B$ variables come in the algebra (\ref{3.16}) on the same
footing and play the role of magnetic and electric fields correspondingly.
This duality seems to generalize the well-known Kramers-Wigner duality and
implies that there may exist a relation between the strong- and
weak-coupling expansions.

In this paper we considered only the pure \Y theory without matter fields.
It would be very interesting to include fermions in the consideration.

The Poisson algebra (\ref{3.16}) can be easily quantized and one gets the
quadratic algebra which was introduced in \cite{ags} to quantize the
Chern-Simons theory. The classical $r$-matrices $\rb$ are to be replaced by
the $R$-matrices $R_{\pm}(q)=1+i\hbar\g\rb +\cdots$, where $q={\rm
e}^{i\hbar\g}$. It is of no problem to check that in quantum theory the \G
are first-class constraints and one can construct  quantum
Hamiltonians which commute with the \G. However the representation theory
of the quantized algebra is at present unknown.

Let us note that $q$ has a
nonpolynomial dependence on the Planck constant $\hbar$ and thus already
"tree" correlation functions of the models will have a nonpolynomial
dependence on $\hbar$ as well. It seems to be an indication that
correlation functions of the models correspond to a summation over
infinitely-many number of the usual Feynman diagrams. It is not excluded
that the parameter $\g$ plays the role of an infrared cut-off. Due to the
fact that there is the additional parameter $\g$ for the models one may
expect that these models have more rich phase structure than the usual
lattice gauge theory.

Let us finally notice that $q$-deformed lattice gauge theory was considered in
refs.\cite{fock,b,ags,br} in connection with  the Chern-Simons theory.

{\bf Acknowledgements:} The author would like to thank A.Alekseev,\break
G.Arutyunov, A.Gorsky, V.Rubtsov  and  A.A.Slavnov for discussions. He is
grateful to Professor J.Wess for kind hospitality and the Alexander von
Humboldt Foundation for the support.  This work has been supported in part
by ISF-grant MNB000 and by the Russian Basic Research Fund under grant
number 94-01-00300a.

\end{document}